# Laser ablation and fragmentation of Boron in liquids

K.O. Aiyyzhy[1], E.V. Barmina[1], V.V. Voronov[1], G.A. Shafeev[1,2], G. G. Novikov[2], O.V. Uvarov[1]

1Prokhorov General Physics Institute of the Russian Academy of Sciences, Vavilova 38, 119333 Moscow, Russia

[2]National Research Nuclear University MEPhI (Moscow Engineering Physics Institute), Kashirskoe sh. 31, 115409 Moscow, Russia

## Abstract

Nanoparticles of elemental Boron are generated for the first time by laser ablation of a sintered Boron target in liquid isopropanol and subsequent laser fragmentation of the suspension. For this purpose an ytterbium doped fiber laser was used at wavelength of 1060-1070 nm, pulse repetition rate of 20 kHz, and pulse duration of 200 ns. The size of Boron nanoparticles after ablation and fragmentation of the suspension is around 30 nm. Nanoparticles are made mostly of Boron and carbon, some particles have carbon shell. Allotropic composition of nanoparticles differs from that of the initial Boron target. Possible applications are discussed.

## Introduction

Recently, there has been a great interest in studying the combustion of suspensions of boron nanoparticles in hydrocarbons. Boron is superior to hydrocarbons in terms of volumetric heat of combustion and in terms of specific mass heat of combustion [1]. These qualities, along with the low toxicity of boron nanoparticles and the possibility of their industrial production, make it possible to consider suspensions of boron nanoparticles in hydrocarbons as promising energy-intensive composite fuels. Studies show that the use of boron nanoparticles as an additive in composite fuel gives a significant increase in temperature in the combustion flame, which affects the energy intensity of the fuel [2, 3].

Also, recent studies show that the use of boron nanoparticles in neutron capture therapy (BNCT) is a promising direction in radiotherapy [4]. BNCT is the selective destruction of tumor cells by accumulating a stable boron-10 isotope in them and subsequent irradiation with neutrons. As a result of the absorption of a neutron by boron, a nuclear reaction occurs with the release of energy in the cell, which leads to its death. The main advantage of boron neutron capture therapy is the selective destruction of malignant neoplasms, in which healthy tissues are not injured. Currently, boron-containing chemical compounds are used for BNCT procedures. The main disadvantage of these compounds is the relatively low content of boron and the chemical method of obtaining these particles. This method of obtaining is accompanied by the formation of additional impurities, which makes it impossible to use in biomedicine. In the present work, it is proposed to use the technique of laser ablation of solids in liquids.

Laser ablation of solids in liquids is well-established technique that allows generation of large variety of nanoparticles [5]. It consists in laser irradiation of bulk solids immersed into the liquid transparent at laser wavelength (see the comprehensive review [6] and references herein). The parameters of generated nanoparticles (size distribution, chemical composition, etc.) depend on number of experimental parameters, such as laser wavelength, laser pulse duration, laser fluence on the target and laser peak intensity, nature of the liquid. Laser fragmentation is well-known process [7-10]. It may be considered as a modification of laser ablation in liquids. It consists in laser exposure of either micro- or nano-powder suspended in a liquid without bulk target. Laser radiation interacts with particles that are inside the laser beam waist in the suspension during the laser pulse. As a role, the fragmentation proceeds through melting of the particles and their interaction with surrounding vapors of the liquid. To reduce size of NPs laser fragmentation method was proposed in [7]. As it can be seen using of laser fragmentation method allows changing the size of NPs and their morphology. Further this process has been thoroughly studied both experimentally and theoretically [8-9]. Laser fragmentation is particularly suitable for generation of nanoparticles from micro-powders of some solids [10].

Surprisingly, the technique of laser ablation in liquids has never been used for generation of Boron nanoparticles, to the best of our knowledge. The goal of this work is to apply the processes of laser ablation and fragmentation for obtaining boron nanoparticles using different ablation schemes.

**Experimental**

In this work boron nanoparticles were obtained by laser ablation of solid Boron target in isopropanol in a continuous-flow cell. A typical layout of the experimental setup is shown in Fig. 1.

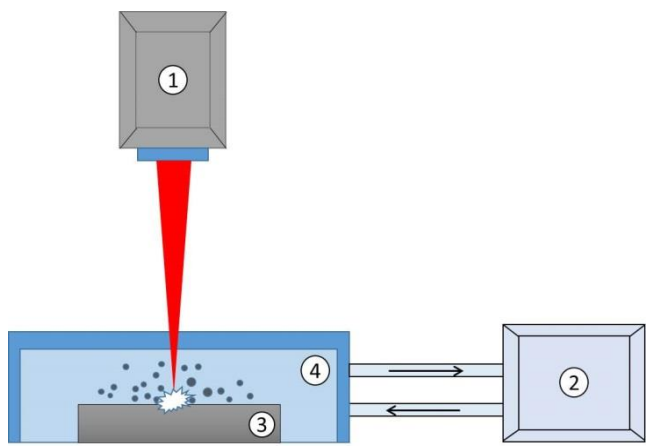

Fig. 1. Schematic of the experiment on fabrication of B nanoparticles in a continuous-flow cell: laser (*1*), circulation pump (*2*), B target (*3*), isopropanol (*4*). The cell walls are cooled with flowing water (not shown).

A piece of sintered boron was used as the target, and isopropanol was the working fluid. The choice of liquid was justified by its low reactivity with boron. In the first experiments, water was used as the liquid. However, during the laser interaction of laser radiation with matter in this case, the formation of boranes (borohydrides) was observed, which instantly volatilized into the environment and burned. The choice of the flow cell was due to the high productivity of the nanoparticles production process. The source of laser radiation was an ytterbium doped fiber laser with an average power of 20 W, a wavelength 1060–1070 nm, a pulse repetition rate of 20 kHz, pulse energy of 1 mJ, pulse duration of 200 ns. The lens focal length was 207 mm.

During laser irradiation of a bulk sintered target, the process of ablation has the character of spallation. The formation of individual micro-fragments was observed due to mechanical damage of the target under laser beam. Similar spallation of the target was previously observed in case of laser ablation of Selenium target in water [11]. This did not allow the creation of stable colloids with nanometer-sized particles. Therefore, to reduce the size of nanoparticles, the colloidal solution of generated boron nanoparticles was subjected to laser fragmentation. In this case, boron micro-powder suspensions obtained by laser ablation of the solid Boron target was irradiated with a laser beam entering through the glass window of the cell from below (Fig. 2.). This scheme makes it possible to control the size distribution of nanoparticles by varying the depth of focusing into the solution and the fragmentation time.

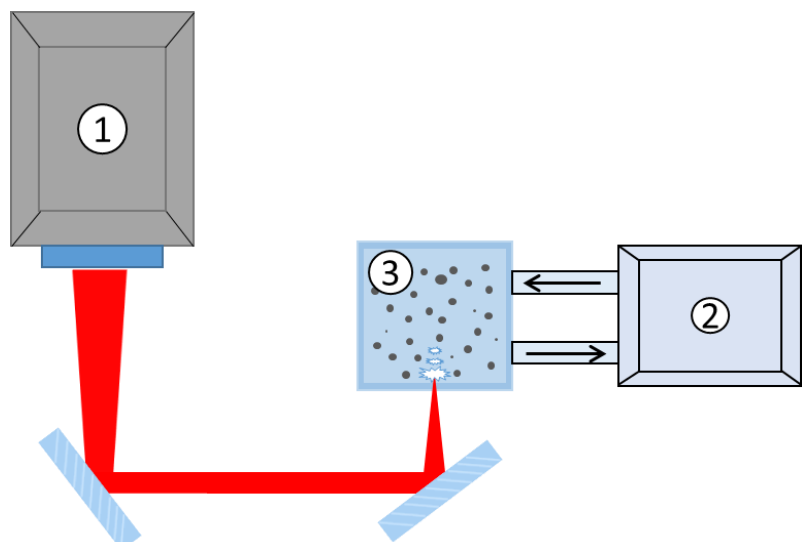

Fig. 2. Schematic of the experiment on fragmentation of B nanoparticles: laser (*1*), circulation pump (*2*), colloidal solution of B nanoparticles (*3*). The cell walls are cooled with flowing water (not shown).

The extinction spectra of colloidal solutions of boron nanoparticles in the optical range (250–800 nm) were measured using an Ocean Optics spectrometer.

The study of the morphology of the obtained nanoparticles was carried out with Carl Zeiss transmission electron microscope (TEM) with an accelerating voltage of 200 kV.

The behavior of the size distribution function of nanoparticles with the time of laser fragmentation was studied using a measuring disk centrifuge CPS Instruments 24000.

The diffraction patterns were recorded on a Bruker D8 Discover A25 DaVinsi Design X-ray diffractometer.

**Results and discussions**

**Laser ablation**

The extinction spectrum of Boron nanoparticles is characterized by weak optical features in the vicinity of 600 nm. This is due to the value of bandgap of this material, which varies between 2.0 and 2.1 eV depending on the crystallographic phase [12]. The decrease in the nanoparticles production rate during irradiation can be estimated from the change in the optical density at a wavelength of 600 nm (Fig. 3). The optical density of the solution increases with ablation time but at various rates. This is due to the absorption of the laser radiation in already formed colloidal solution. As the result, the laser fluence on the target gradually decreases and the ablation rate drops down.

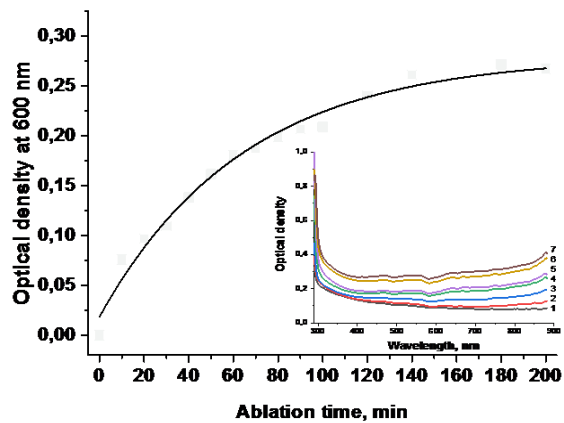

Fig 3. Dependence of the optical density of a colloidal solution of B nanoparticles at a wavelength of 600 nm on the laser ablation time. The inset shows the extinction spectra for various ablation times: 10 (1), 20 (2), 40 (3), 80 (4), 100 (5), 120 (6) and 200 min (7).

The size distribution of the particles formed during laser ablation of a boron target in a flow cell (Fig. 4) is bimodal. The fraction of small (less than 100 nm) nanoparticles is rather small, while the fraction of larger particles is still relatively high.

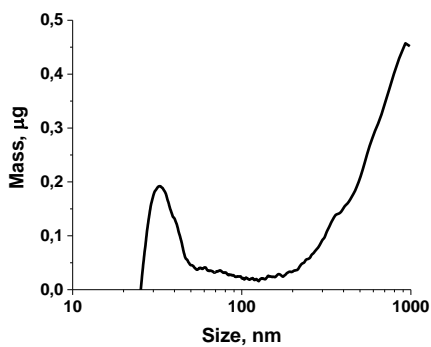

Fig 4. Size distribution of B nanoparticles obtained using a disc centrifuge CPS 24000. The colloidal solution of boron nanoparticles was obtained by ablation in isopropanol in a continuous-flow cell using an ytterbium-doped fiber laser for 200 min.

Large micro-particles are formed in this case because of the high laser fluence on the target. The reason is probably that the boron target undergoes intense spallation.

The X-ray diffractograms of the initial target and of generated nanoparticles are presented in Fig. 5. The diffractograms of the initial target (Fig. 5, a) is rather complicated but perfectly matches the crystalline structure made of 314 units of $B_7$ cells according database Powder Diffraction File-2, version 2011. This crystallographic structure is formed under sintering of Boron powder at elevated temperature [13]. However, the crystallographic structure of generated nanoparticles is different. It can be seen that ablated NPs undergo allotropic modification. They are made of rhombohedral Boron and of elementary units of centrosymmetric $B_7$ (Fig. 5, b).

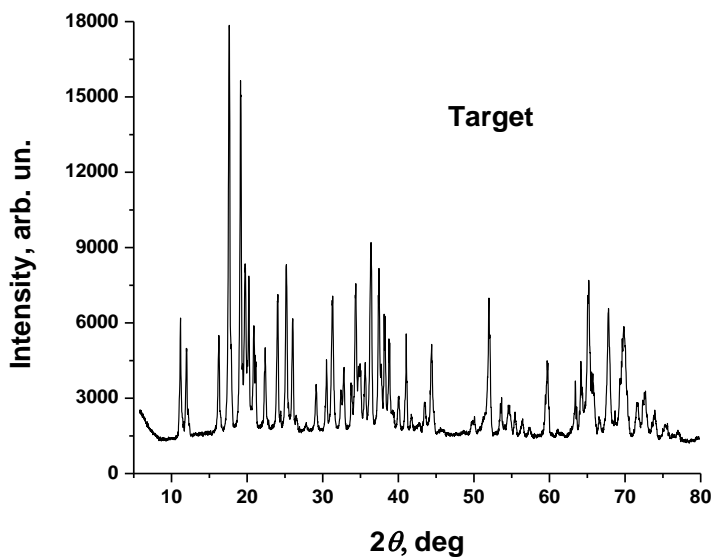

a

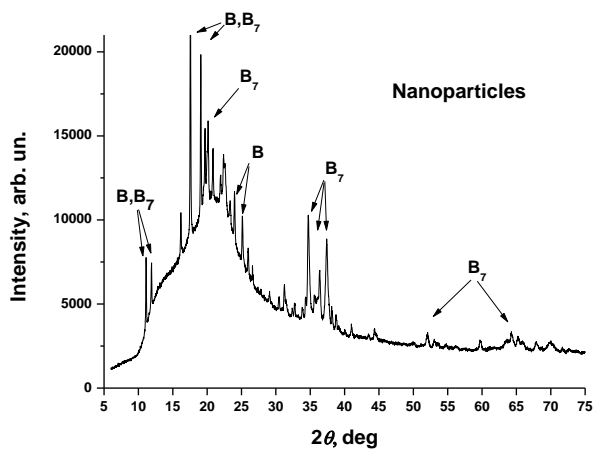

b

Fig 5. X-ray diffraction patterns of the initial sintered Boron target (a) and nanoparticles generated by laser ablation of this target (b).

Allotropic modification of Boron as the result of laser ablation is rather typical for this element and its compounds. Similar modification was previously observed in case of laser ablation of $Fe_2B$ target enriched in [10]B content in isopropanol [14]. Apart from nanoparticles of $Fe_2B$ itself there are also nanoparticles of Fe and molecular species of $B_{12}$ and $B_{28}$, which are absent in the initial target.

TEM studies show that the sample after laser ablation is highly inhomogeneous. The size of some particles lies in sub-micrometer range. Also, the elemental composition of some particles is very different from point to point. Boron and carbon contents may vary from point to point from 85 to 15%. Moreover, some parts of the samples are modified under the electron beam of the microscope to pure carbon. Therefore, the main elements are boron and carbon. However, X-ray diffractogram of nanoparticles (see Fig. 5, b) does not contain carbon. Therefore, its detection by TEM should be considered as an artifact. Most probably, this carbon is related to the products of isopropanol decomposition during laser ablation, and these products are decomposed under the electron beam of the microscope.

Finally, the liquid isopropanol undergoes changes during the laser ablation of the Boron target. These changes are caused by high temperature during the ablation. Also they are due to the light emission of the plasma on the target surface and by plasma itself. Evolution of the optical density of the colloidal solution of Boron nanoparticles with the time of laser ablation is shown in Fig. 6.

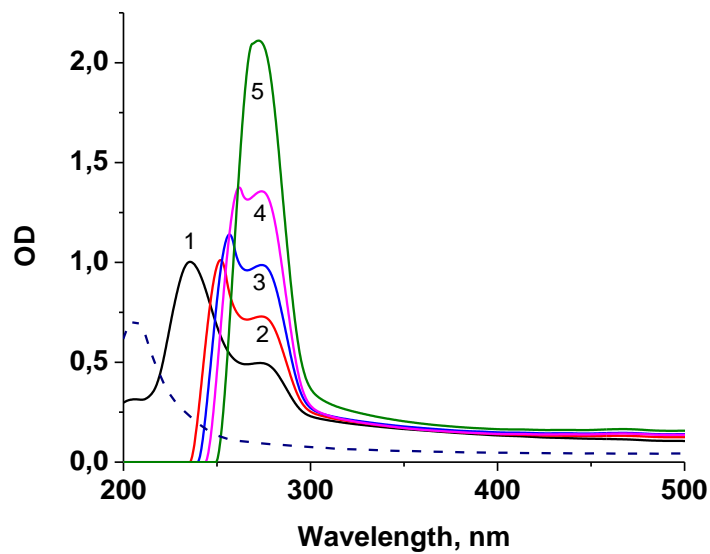

Fig. 6. Evolution of the extinction spectra of the colloidal solution in UV range with time of laser ablation of Boron target. Time of laser ablation: 1 -10, 2-30, 3 - 50, 4-70, and 5 - 90 minutes. Spectra are recorded with pure isopropanol as a reference. Dashed line is isopropanol spectrum against empty quartz cell.

The main contribution to the extinction of the solution between 300 and 500 nm give nanoparticles of Boron. UV range of the extinction spectrum is attributed to the hydrocarbons. New peak at 272 nm arises with the increase of laser ablation time. This new peak is attributed to the product of isopropanol decomposition. Note that pure isopropanol is characterized by another extinction spectrum shown in Fig. 6 by dashed line. The modification of hydrocarbons under laser ablation is well known. The laser ablation in hydrocarbons is accompanied by emission of molecular hydrogen and modification of the hydrocarbon to wide range of other hydrocarbons [15]. Finally, optical properties of decomposed isopropanol under laser beam do not affect absorption spectrum of Boron nanoparticles.

Summarizing the above results, to generate small boron nanoparticles colloidal solution of boron nanoparticles laser fragmentation method should be used. In this case, the suspension of boron micro- and nanoparticles generated by laser ablation was exposed to laser beam in the colloidal solution in absence of the target.

**Laser fragmentation**

The changes in the nanoparticles size distribution during the colloidal solution irradiation (Fig. 7) show that before the irradiation the main mass of the nanoparticles ensemble is located in large particles.

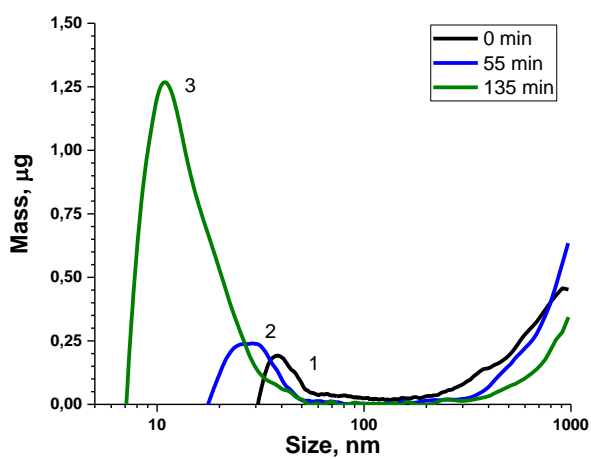

Fig 7. Mass distributions on size of B nanoparticles for different fragmentation times: 0 (1), 55 (2), and 135 min (3). The volume of the colloid portion in the centrifuge is of 100 μl.

The mass of small nanoparticles increases with fragmentation time. After sufficiently long irradiation, particles less than 30 nm in size dominate (the distribution has a peak at a size of 15 nm). However, not all large particles are fragmented in the final colloidal solution. Apparently, complete fragmentation requires even longer irradiation times.

STEM view of nanoparticles after fragmentation is shown in Fig. 8.

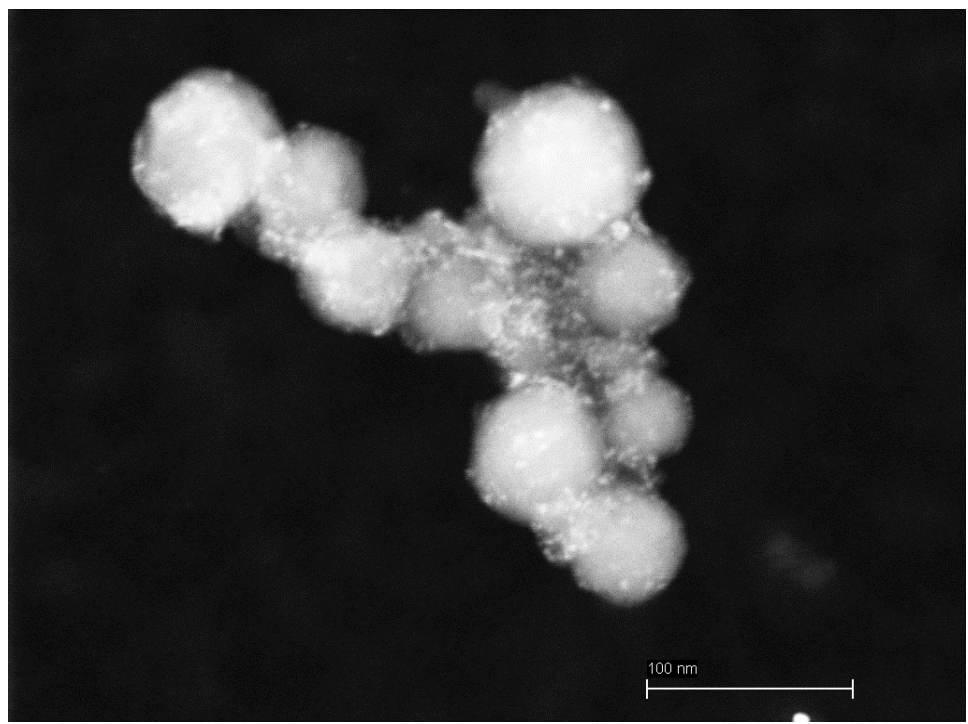

Fig. 8. View of nanoparticles after fragmentation in scattered electrons (STEM). Scale bar denotes 100 nm.

One can see that nanoparticles with average size 50 nm are covered by diffuse layers of another heavier material, most probably by carbon.

TEM and STEM views of Boron nanoparticles after laser fragmentation are presented in Fig. 9.

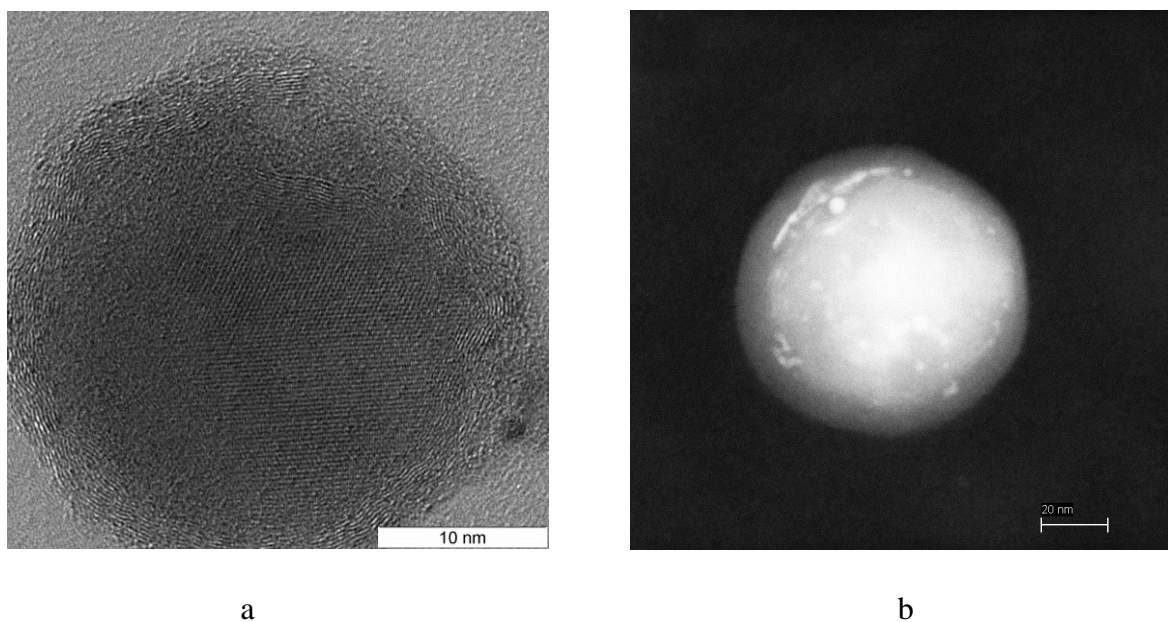

a                                                        b

Fig. 9. View of individual Boron nanoparticle in transmission mode (a) and in scattered electrons STEM (b). Scale bars denote 10 (a) and 20 nm (b).

A shell around spherical particle is clearly visible in transmission and in scattered electrons. Crystallographic planes of the shell are visible in Fig. 9, a. The planes are also projected on the central part of the particle.

Tracing of the elemental composition of the particle across its diameter indicates that carbon content is more or less homogeneous in the central part of the particle and increases on its edges. Therefore, this carbon shell covers the whole particle.

The nanoparticles contain Boron and carbon, similarly to nanoparticles prepared by laser ablation of Boron target. Mapping the elements in isolated nanoparticle is shown in Fig. 10. Boron is located mostly in the central part of the particle, while carbon content is more pronounced on its periphery and may correspond to the carbon shell. There are also some traces of oxygen.

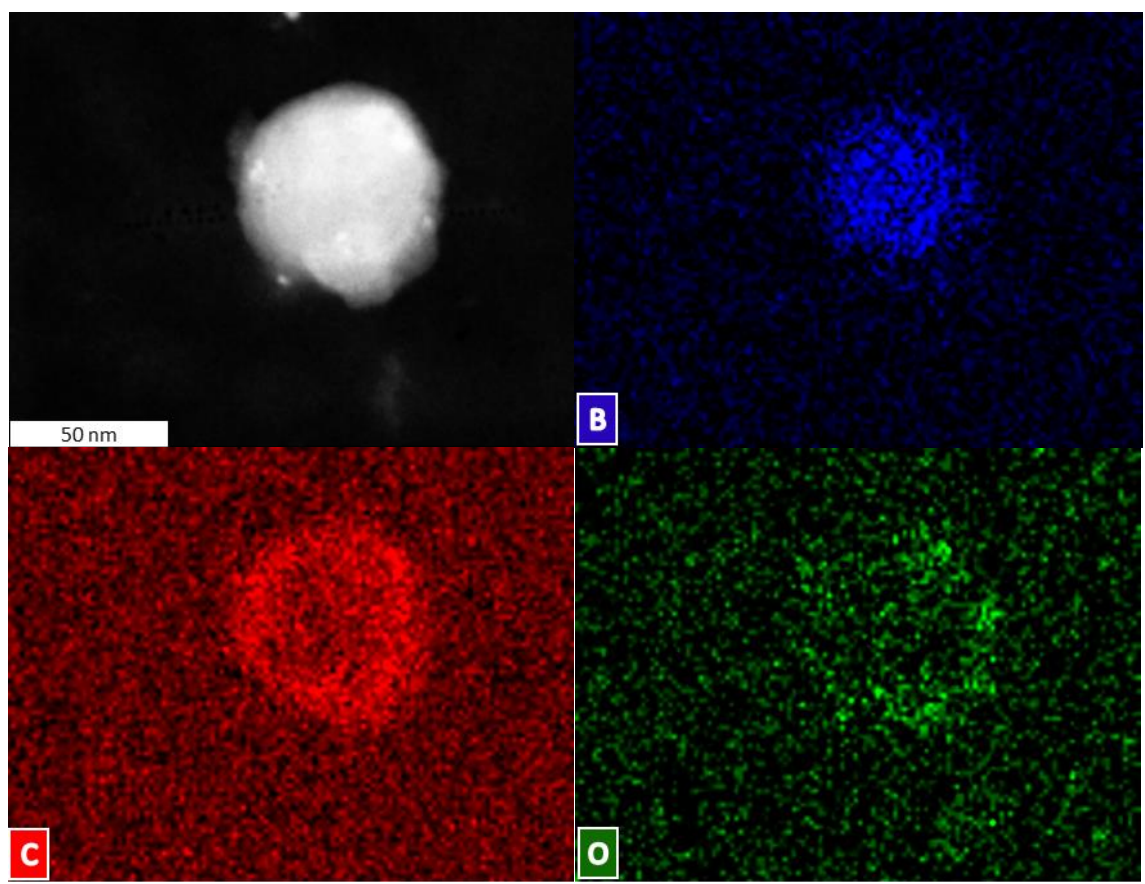

Fig. 10. TEM view of individual nanoparticle and distributions in it of Boron (B), carbon (C), and oxygen (O). Scale bar denotes 50 nm.

XRD diffraction pattern of fragmented Boron nanoparticles is virtually identical to that of nanoparticles obtained by laser ablation of Boron target (see Fig. 5).

Electron energy loss spectroscopy (EELS, not shown here) indicates that the peak of Boron is splitted onto two peaks. The first peak corresponds to Boron itself, while the second is situated for 7 eV lower in energy. This may be attributed to the formation of some Boron-containing compound; most probably this is Boron carbide. This point requires further investigations by other methods, e.g., by Raman spectroscopy.

**Conclusion**

Thus, nanoparticles of elemental Boron have been successfully synthesized using the technique of laser ablation of a sintered Boron target in liquid isopropanol. Laser ablation of the target results in the formation of suspension of relatively large sub-micrometer Boron particles. Further laser fragmentation of this suspension leads to formation of smaller nanoparticles with average size of 30 nm. The nanoparticles contain Boron, carbon, and some traces of oxygen. TEM studies show that small nanoparticles of Boron possess carbon-containing shell. The boron nanoparticles obtained in these experiments will be investigated as additives to composite hydrocarbon fuel. The presence of carbon shell on Boron nanoparticles gives the advantage for nanocomposite hydrocarbon fuels since carbon may serve as the ignitor of Boron burning. The use of generated nanoparticles in BNCT process is also of high interest.

The work was supported by Russian Science Foundation, Grant № 20-19-00419. We are grateful to the Centre of Collective Use of the Prokhorov General Physics Institute of the Russian Academy of Sciences for the TEM images of nanoparticles and X-ray diffraction data.

**References**

1. G. Young, K. Sullivan, M. R. Zachariah, K. Yu, Combustion characteristics of boron nanoparticles, Combustion and Flame 156 (2009) 322–333. https://doi.org/10.1016/j.combustflame.2008.10.007

2. S. Karmakar, S. Acharya, K. M. Dooley, Ignition and Combustion of Boron Nanoparticles in Ethanol Spray Flame, Journal of Propulsion and Power 28 (2013) 707-718. https://doi.org/10.2514/1.B34358

3. E.V. Barmina, M.I. Zhilnikova, K.O. Aiyyzhy, V.D. Kobtsev, D.N. Kozlov, S.A. Kostritsa, S.N. Orlov, A.M. Saveliev, V.V. Smirnov, N.S. Titova, G.A. Shafeev (2022), Experimental investigation of diffuse burning of suspension of Boron nanoparticles in isopropanol, Doklady Physics, 67 No. 2 (2022) 39–43.

4. M. Dymova, S. Taskaev, V. Richter, E. Kuligina, Boron neutron capture therapy: current status and future perspectives, Canc. Com. 40 (2020) 406-421. https://doi.org/10.1002/cac2.12089

5. P. V. Kazakevich, A. V. Simakin, V. V. Voronov, G. A. Shafeev, Laser-induced synthesis of nanoparticles in liquids, Applied Surface Science 252 (2006) 4373-4380. https://doi.org/10.1016/j.apsusc.2005.06.059

6. D. Zhang, B. Gökce, S. Barcikowski, Laser Synthesis and Processing of Colloids: Fundamentals and Applications, Chemical Reviews 117 (2017) 3990-4103. https://doi.org/10.1021/acs.chemrev.6b00468

7. M. Procházka, P. Mojzeš, J. Štěpánek, B. Vlčková, P. Y. Turpin, Probing applications of laser ablated Ag colloids in SERS spectroscopy: improvement of ablation procedure and SERS spectral testing, Analytical Chemistry 69 (1997) 5103-5108. https://doi.org/10.1021/ac970683+

8. L. Delfour, T. E. Itina, Mechanisms of ultrashort laser-induced fragmentation of metal nanoparticles in liquids: numerical insights, The Journal of Physical Chemistry C 119 (2015) 13893. https://doi.org/10.1021/acs.jpcc.5b02084

9. Ziefuss A. R. et al., Laser fragmentation of colloidal gold nanoparticles with high-intensity nanosecond pulses is driven by a single-step fragmentation mechanism with a defined educt particle-size threshold, The Journal of Physical Chemistry C (2018) 122, 22125.

10. M. Zhilnikova, E. Barmina, I. Pavlov, A. Vasiliev, G. Shafeev, Laser fragmentation of $Ag_2O$ micropowder in water, Journal of Physics and Chemistry of Solids160 (2022) 110356. https://doi.org/10.1016/j.jpcs.2021.110356

11. K. O. Ayyyzhy, V. V. Voronov, S. V. Gudkov, I. I. Rakov, A. V. Simakin, and G. A. Shafeev, Laser Fabrication and Fragmentation of Selenium Nanoparticles in Aqueous Media, Physics of Wave Phenomena Vol. 27 No. 2 (2019) 113–118. DOI: 10.3103/S1541308X19020055

12. E. Yu. Zarechnaya; L. Dubrovinsky; N. Dubrovinskaia; Y. Filinchuk; D. Chernyshov, V. Dmitriev; N. Miyajima, A. El Goresy, et al. Superhard Semiconducting Optically Transparent High Pressure Phase of Boron. Phys. Rev. Lett. 102-18 (2009)185501. DOI: 10.1103/PhysRevLett.102.185501

13. B. Callmer, An accurate refinement of the β-rhombohedral Boron structure, Acta cryst B33 (1977) 1951-1954. https://doi.org/10.1107/S0567740877007389

14. E. V. Barmina, I. N. Zavestovskaya, A. I. Kasatova, D. S. Petrunya, O. V. Uvarov, V. V. Saraykin, M. I. Zhilnikova, V. V. Voronov, G. A. Shafeev, S. Yu. Taskaev, Laser ablation of $Fe_2B$ target enriched in [10]B content for boron neutron capture therapy, www.arxiv.org 2109.03608.

15. I.V. Baymler, E.V. Barmina, A.V. Simakin, G.A. Shafeev, Generation of hydrogen under laser irradiation of organic liquids, Quantum Electronics **48** (8) (2018) 738 – 742. https://doi.org/10.1070/QEL16648